\documentclass[final]{ias2}

\usepackage{graphicx} 
\usepackage{multirow}
\usepackage{array}

\usepackage{hyperref} 

\usepackage{amsmath,amssymb}

\newcommand{\bs}[1]{\boldsymbol{#1}}

\newcommand{\ket}[1]{\left|#1\right>}
\newcommand{\cs}{c_{\rm s}}
\newcommand{\kB}{k_{\textsc{b}}}
\newcommand{\ud}{\textrm{d}}
\DeclareMathOperator{\Li}{Li}

\begin{document}

\markboth{Chance and Chandra}{Goodman et al.}

\title{Chance and Chandra}

\author[pct]{Jeremy Goodman} 
\email{jeremy@astro.princeton.edu}
\author[pct]{Zachary Slepian} 
\email{zslepian@princeton.edu}
\address[pct]{Princeton University Observatory, Princeton, NJ, USA}

\begin{abstract}
  A few examples are given of Chandra's work on statistical and stochastic problems that
  relate to open questions in astrophysics, in particular his theory of dynamical
  relaxation in systems with inverse-square interparticle forces.  The roles of chaos and
  integrability in this theory require clarification, especially for systems having a
  dominant central mass.  After this prelude, a hypothetical form of bosonic dark matter
  with a simple but nontrivial statistical mechanics is discussed.  This makes for a
  number of eminently falsifiable predictions, including some exotic consequences for
  dynamical friction.
\end{abstract}


\pacs{Appropriate pacs here}
 
\maketitle


\section{Introduction}

Probabilistic and statistical considerations dominated several of Chandrasekhar's
contributions to astronomy and astrophysics.  Among his less-well-remembered contributions
to astronomy proper, as opposed to astrophysics, are a series of papers with M\"unch on
brightness fluctuations in the star fields of the Milky Way \cite[beginning with][]{CM50}
and one on the inference of the distribution of stellar spins from observed values of
$v\sin i$ \cite{CM_spin}.  In a paper that deserves to be better remembered
\cite{C_scint}, Chandra introduced the use of structure functions to describe the
refractive-index fluctuations that degrade ground-based astronomical images (``seeing'').
These methods have been used ever since by students of atmospheric and interstellar
scintillation, although usually without recognizing the debt owed to Chandrasekhar.
Better known is Chandra's magisterial review article on random walks and other stochastic
processes \cite{C_RMP}.

The problems of this type for which Chandra is most famous are those of dynamical
relaxation and dynamical friction.  He was not solely responsible for these
developments---the problem goes back at least as far as Jeans \cite{Jeans}, and a fairly
complete solution for the velocity diffusion coefficients of a plasma subject to Coulomb
encounters was given in the 1930s by Landau \cite{Landau36}, but Chandra's subsequent
emphasis on the probabilistic aspects was distinctive.  He considered the probability
distribution of the gravitational force experienced at a given point in space due to an
infinite field of point masses whose locations form a Poisson process (the Holtzmark
distribution), for which $\mbox{Prob}(>f)\propto f^{-5/2}$ at large values of the
magnitude $f$ of the force, and he then considered the effect of the motion of these
gravitating masses on a test mass \cite{C41}.  Rather than describe these as a series of
impulsive two-body encounters, as done in most textbooks \citep[e.g.][]{BT} and
indeed in his own monograph \cite{C_PSD}, Chandra treated the summed gravitational forces
of the field particles as a time-dependent stochastic process.  In Chandra's hands, this
approach described the diffusion of the test particle's velocity, but not (at least in the
first instance) the back-reaction on the field particles; therefore it predicted
indefinite increase in the test particle's kinetic energy.  Chandra recognized the need
for a drag term---dynamical friction---to preserve a Maxwellian equilibrium distribution
\cite{C_DF}.

It is well known that the usual treatments encounter a logarithmic divergence, the usual
cure for which is to estimate lower and upper bounds for the impact parameters of two-body
encounters.  The divergence at small impact parameters is straightforward to remove by
replacing the impulse approximation with exact formulae for hyperbolic close encounters,
as Chandra demonstrated, so that it suffices in most cases to set $b_{\min}\sim
G(m_{\rm t}+m_{\rm f})/\langle v_{\rm rel}^2\rangle$, these symbols having the usual meanings.  But
the divergence at large impact parameters is more problematic.  It is conventional to
remark that the treatment of the field particles as statistically uniform breaks down
at distances comparable to the size or density scale height of a self-gravitating system
($R$) and therefore to cut off the impact parameter at $b_{\max}\sim R$, so that (since
$\langle v_{rel}^2\rangle\sim Gm_{\rm f}N/R$) the argument of the logarithm is of order $N$, the
total number of particles.  Chandra, however, argued for a cutoff at the mean
interparticle distance, $D\sim R/N^{1/3}$.  This choice seems to be in some tension with
his view of the gravitational field as a collective process.

Although it is now accepted that $b_{\max}/b_{\min}\propto N$ in self-gravitating systems,
much ink has been shed over the proper choice of the constant of proportionality.  As a
practical matter, it makes little difference for large $N$.  Moreover, having worked in
stellar dynamics for his Ph.D. and several years thereafter, the speaker and first author
of this talk (JG) takes the view that the whole discussion is futile because there can be
no such thing as a universally correct choice.  The logarithm must depend at some level
upon the details of the density profile of the system and upon where the test particle
orbits within it.  Furthermore, the root of the difficulty with large impact parameters
lies in the insistence upon Markovian approximation: that is, the assumption that the
probability distributions or moments of the test particle's velocity and position can be
described by differential equations.  This may be a very good approximation for gas
particles interacting via short-range collisions, or even for Coulomb encounters in a
plasma, since they are naturally cut off at the Debye length.  But gravitational
encounters have a duration that increases, on average, in proportion to impact parameter,
$b$.  Even in an infinitely system of field particles, the evolution of the probability
distribution $P[\bs{v},t|\bs{v}_0,t_0]$ for the test particle's velocity, should be
insensitive to encounters more distant than $b(t,t_0)\equiv |t-t_0|\langle v_{\rm
  rel}^2\rangle^{1/2}$; stars more distant than this exert a net acceleration that is
effectively constant during the interval $t-t_0$.  The divergence of the net acceleration
as the size of the system tends to infinity is removable (by shifting to a freely-falling
frame) because it is constant for all finite time intervals.  To reformulate dynamical
relaxation along these lines would require integral rather than differential equations in
time, however.  The slight theoretical ambiguity represented by the usual logarithm is a
price well worth paying for convenience in practical applications.

There is another bit of unfinished business regarding dynamical friction that is probably
more urgent, especially because of the discovery of extrasolar planetary systems and the
consequent revival of interest in celestial mechanics.  (Chandra did not work in celestial
mechanics.  Perhaps it seemed to him to be a closed subject.  But in view of the
mathematical elegance of the subject and the new questions brought to light by computer
calculations as well as planetary discoveries, it is likely that he would take interest if
he were alive today.)  This unfinished business concerns the relationship between
dynamical relaxation and chaos.  Chandra's analysis presumes that two given particles
encounter one another just once since he treats their unperturbed trajectories as
rectilinear.  This eliminates the possibility of resonances, since resonances require
repeated encounters.  Celestial mechanics, however, is rife with resonance; and chaos
begins where resonances overlap \cite{Chirikov}.  It seems that one uses the language of
celestial mechanics when the number of gravitating bodies is small, $N\lesssim 10$, but
that of dynamical relaxation when $N\gtrsim 10^3$.  Ambiguous intermediate cases exist:
For example, the minimum number of particles in the core of a cluster of equal point
masses undergoing core collapse (and rebound) is thought to be $\lesssim 30$ \cite{HH03};
and the number of significant bodies in a young planetary system might
well lie in the range 10-100 before this number is reduced by consolidation or ejection.
It would be interesting, and perhaps useful for application to such intermediate cases, to
understand these two domains of relaxation and resonant interaction as different limits of
a unified statistical theory of Nbody systems.  Some steps in this direction have been
taken \cite{RT,TW}.

After this prelude, the remainder of this talks is devoted to a rather different topic: a
(probably counterfactual) species of interacting dark matter.  The connection to Chandra's
work is admittedly tenuous, and consists mainly in the unusual properties of dynamical
relaxation for such a dark matter.

\section{Dark matter and its possible collisionality}

Nonbaryonic dark matter (hereafter DM) makes up some $22\%$ of the critical density
$3H_0^2/8\pi G$ \cite{WMAP}.  Notwithstanding experimental searches and suggested evidence
for annihilation signals in cosmic backgrounds, DM is so far known only through its
gravitational effects, including galactic rotation curves, virial temperatures of X-ray
emitting gas in clusters of galaxies, and the power spectrum of temperature fluctuations
of the cosmic microwave background.  In compliance with Occam's Razor, DM is therefore
usually presumed to have a minimum of non-gravitational interactions.  However, were he
alive today, William of Occam would presumably allow DM to have a minimum of additional
properties if those would resolve important problems in particle physics as well as
astrophysics and cosmology.  In this spirit, the axion and supersymmetric weakly
interacting massive particles (WIMPs) have been put forward as dark-matter candidates.
The existence of axions might explain CP conservation in the strong interactions
\cite{PQ77}, while supersymmetry would unify fermions and bosons, and perhaps also the
strong, weak, and electromagnetic coupling constants at high energies \cite{DRW}.  WIMPs
are favored by the consideration that if they were produced thermally in the early
universe, then the present-day density of DM is consistent with WIMP masses in the range
1-100 GeV and annihilation cross-sections comparable to what would be expected from the
weak interactions \cite{JKG}.

Effectively non-interacting DM would be consistent with most of what is known or believed
about the present universe, and so it is conventional to treat DM as collisionless.  This
assumption lends itself to predicting the dynamics of DM via Nbody simulations.
Occasionally, however, attention is drawn to apparent discrepancies between collisionless,
dynamically ``cold'' DM and certain features of galactic structure on small scales.  It is
suggested on this basis that the DM may have additional properties such as an appreciable
initial velocity dispersion (``warm'' DM), or even some collisionality.  One such
discrepancy is the dearth of small satellite galaxies compared to the abundance of DM
subhalos predicted to lie within the DM halos of luminous ($\gtrsim L_*$) galaxies like
the Milky Way \cite{Klypin1999}.  This, however, might be explained by a low efficiency of
star formation in the subhaloes \cite{SWTS}.  Another discrepancy is the prevalence and
persistence of apparently rapidly rotating galactic bars despite the dynamical friction
exerted on these bars by the DM halo\cite{DS2000}.  But perhaps the dynamical friction is
reduced by resonant effects too delicate for standard Nbody simulation methods
\cite{WK2007}.  A decade ago, the most talked-about discrepancy was an apparent minimum
core radius for dark halos inferred from rotation curves of dwarf galaxies
\cite{Moore1994,Moore2000,vdB2000,Swaters2009}.  To solve these apparent problems, Spergel
\& Steinhardt \cite{SS2000} suggested that the (WIMPy) DM might have a collisional
mean-free-path in the range $1\mbox{ kpc}\lesssim\lambda_{\rm mfp}\lesssim 1\mbox{ Mpc}$.
This would efficiently erode DM subhalos and cause cuspy DM cores with profiles
$\rho_{\textsc{dm}}\propto r^{-\alpha}$ in the range $0<\alpha<2$ to be smoothed out, at
least initially, because the dynamical temperature associated with such profiles increases
with radius.  However, as the authors acknowledged, the same process of thermal conduction
in a self-gravitating system leads eventually to core collapse, i.e. an indefinite
\emph{increase} in central density.  They tried to argue that the timescale for collapse
exceeds a Hubble time.  But the timescale for core collapse cannot greatly exceed that for
cusp smoothing except in carefully engineered profiles that would not prevail generally.

Galaxy clusters in collision provide perhaps the most decisive limits on self-interacting
DM.  Comparison of the distributions of gas (traced by X-rays) and dark matter (traced by
weak gravitational lensing) in 1E~0657-56, the ``Bullet Cluster,'' has been used to place
a limit on the DM scattering cross section per unit mass: $\sigma/m< 1.25\,{\rm
  cm^2\,g^{-1}}$ at 68\% confidence \cite{Randall_etal2008}.  This translates to
$\lambda_{\rm mfp}\gtrsim 0.5\,\mbox{Mpc}$ for the DM in the Solar Neighborhood.

\section{Repulsive dark matter: Overview} 

A different sort of self-interacting DM was proposed at about the same time as \cite{SS2000}: bosonic,
like the axion, but with a repulsive interaction \cite{Peebles2000,G2000}.  Also like the
axion, these particles were supposed to be born in a Bose-Einstein condensate in the early
universe.  The interaction would naturally be short-range, in fact pointlike, if it
corresponded to the non-relativistic limit of a massive scalar field with a
momentum-independent self-interaction term $V(\phi)>0$.  In the condensate, pressure would
vary only with density.  In fact if $V(\phi)=\kappa\phi^4$, with $\kappa$ a dimensionless
coupling, then it was argued by \cite{G2000} that in the non-relativistic limit $P\ll\rho
c^2$, the pressure and density would be related as in an $n=1$ Emden polytrope,
$P=K\rho^2$, with $K=3\kappa/2m^4$ in units $\hbar=c=1$, where $m$ is the particle mass.
As Chandra himself discussed in his monograph on stellar structure \cite{Chandra_SS}, the
cases $n\in\{0,1,5\}$ are the only polytropes for which a (nonsingular) self-gravitating
spherical equilibrium can be found analytically; for $n=1$, the density profile is
$\rho(r)/\rho(0)=\sin(r/a)/(r/a)$, $0\le r\le\pi a$, with scale length $a=\sqrt{K/2\pi G}$.
If all of the DM were still in the condensate, then all nonrotating dark halos in virial
equilibrium would have this size and density profile independent of their masses.  Clearly
this would not be acceptable, especially if the scale length $a\sim 1\mbox{ kpc}$ as
required to match the DM cores of dwarf galaxies.  More complex and extended profiles can
be obtained if the DM has a finite temperature, as discussed below, or where it has not
yet reached thermal or even virial equilibrium.

It was argued by \cite{Peebles2000} and \cite{G2000} that if present-day repulsive dark
matter (hereafter RDM) derives from a relativistic scalar field, then it has acceptable
behavior in the early universe, \emph{i.e.} it does not suppress large-scale structure or
unduly affect primordial nucleosynthesis.  There is some doubt whether even a massive
scalar field has a sensible non-relativistic limit, but we shall not
worry about that here.  We treat the RDM as an assemblage of nonrelativistic point
particles of mass $m$ having a two-body interaction potential whose range is small
compared to the particles' Compton wavelength, so that $U(\bs{x}_1-\bs{x}_2)\to
U_0\delta(\bs{x}_1-\bs{x}_2)$.  The constant $U_0=\tilde U(0)$ is the fourier transform of
$U(\Delta\bs{x})$ at zero momentum.  If the relativistic correspondence were valid and
$V(\phi)=\kappa\phi^4$, then $U_0=3\kappa\hbar^3/m^2c$ at tree level.  We further assume
that that the particles are conserved.  In the relativistic correspondence, this would
require either that $\kappa$ be very small, or else that the particles
be protected from annihilation by some symmetry, e.g. by making $\phi$ a
charged scalar field and its action invariant to global phase changes, since otherwise
they would have an annihilation cross section at $O(\kappa^4/m^2)$ (annihilation of two
particles into one is kinematically forbidden, so the lowest-order graph has two
interaction vertices).

In terms of the two basic parameters of the nonrelativistic model, the core scale length
is $a=\sqrt{U_0/4\pi Gm^2}$, whereas the scattering cross section per unit mass is, to
lowest order, $\sigma/m = mU_0^2/\pi\hbar^4$.  Since these involve independent
combinations of $m$ and $U_0$, the degree of collisionality of RDM is independent of its
minimum core size.  So it seems possible to evade the constraint on $\sigma_{\rm scatt}/m$
set by the Bullet Cluster \cite{Randall_etal2008}.  Also, since $\sigma_{\rm
  ann}/\sigma_{\rm scatt}\propto\kappa^2\propto (m^2 U_0)^2$, reducing the collisionality
reduces the annihilation rate faster.  As will be shown, however, a long mean-free path
leads in this species of DM to unacceptable halo density profiles, at least if collisions
are frequent enough for thermal equilibration over a Hubble time.

\section{Superfluidity, dynamical friction, and galactic bars}

The repulsion makes the condensate, a large number of bosons in the same single-particle
state, a superfluid.  Superfluidity involves a penalty for excitations out of
the degenerate state, which arises from the repulsive interactions.
As a consequence of their indistinguishability, the repulsion between a pair of RDM
bosons in distinct momentum states $\ket{\bs{p}}$ \& $\ket{\bs{p'}}$
[$\left<\bs{p}|\bs{p'}\right>=0$] is greater than it would be if both were in
$\ket{\bs{p}}$ by \citep[e.g.][]{Baym}
\begin{equation}
  \label{eq:gap}
  \Delta E = {\nu}\tilde U(\bs{p}-\bs{p}')+\frac{\bs{p}^2-\bs{p}^{\prime\,2}}{2m}
\approx {\nu}U_0+\frac{\bs{p}^2-\bs{p}^{\prime\,2}}{2m}\,,
\end{equation}
since the exchange energy reduces to ${\nu}\tilde U(0)$
for a short-range potential.  Consequently, a rigid obstacle (``spoon'')
passing through the condensate at relative velocity $\bs{v}$ causes no dissipation if
$|\bs{v}|<v_{\rm crit}\equiv \sqrt{2{\nu}_{\rm c}/m}$ because, viewing the problem
in the rest frame of the spoon, the energy liberated by reducing the kinetic energy of a
particle to zero is less than the exchange energy.  Similarly, two
streams of RDM condensate pass freely through each other if their relative velocity is
less than $\sqrt{2}v_{\rm crit}$ even when the mean-free-path would otherwise be short.
Here ${\nu}_{\rm c}$ is the total number density of the condensate
summed over both streams.

At the time of \cite{G2000}, one of us (JG) thought that the superfluidity of RDM implies
that there should be no dynamical on a moving potential, such as a galactic bar, insofar
as the core of the halo is dominated by the condensate and $v_{\rm bar}<v_{\rm crit}$.
Our current view is somewhat different.  In the limit that the mean-free-path is large
compared to the dimensions of the bar or the core, the original argument is perhaps
correct.  In the opposite limit, while it remains true that the dynamical
friction vanishes below a critical speed $\sim v_{\rm crit}$, the reason is not because
the RDM is a superfluid, but rather because it is an ideal fluid in the classical sense,
i.e. one in which the viscosity based on the mean-free-path and thermal velocity is
negligible.  The important difference between a galactic bar interacting with DM
gravitationally and a spoon interacting with a laboratory fluid is that the gravitational
potential is smooth, whereas the spoon has a surface at which a no-slip boundary condition
applies to the normal (non-superfluid) component.  Thus, even in laminar flow, the spoon
exerts a viscous force on the normal component that vanishes more slowly than linearly
with the viscosity, probably as $Re^{-1/2}$, because there is a laminar boundary layer.
At moderately high Reynolds number, $Re$, the boundary layer on the rear side of the spoon
becomes unstable and triggers turbulence.  For a smooth large-scale potential like that of
a rotating bar, however, there is no such surface and no such boundary layer, and
therefore perhaps no turbulence.  This absence of drag in the limit of small $\lambda_{\rm
  mfp}$ would seem to hold even if the RDM had a substantial normal component, except
insofar as turbulence may arise even without boundary layers in a smooth velocity field at
large $Re$.  

On the other hand, the bar or other moving potential may experience a wave drag if it
couples to a waves whose phase velocity, whether linear or angular, matches that of the
potential itself.  In a pure condensate, below $v_{\rm crit}$, the only significant waves
are sound waves at speed $c_{\rm s}=\sqrt{2K\rho}=v_{\rm crit}/\sqrt{2}$, so that a
linearly moving potential experiences no wave drag if its velocity is subsonic, though a
rotating bar could excite sound waves at large radii were $|\bs{r\times\Omega}_{\rm
  bar}|>c_{\rm s}$.  One can see this explicitly from the quantum-mechanical equation of
motion for the condensate wave function (Gross-Pitaevskii or nonlinear Schr\"odinger
Eqn.):
\begin{equation}
  \label{eq:NLSE}
  i\hbar\frac{\partial\Psi}{\partial
    t}=-\frac{\hbar^2}{2m}\nabla^2\Psi+U_0|\Psi|^2\Psi+(\Phi_{\rm self}+\Phi_{\rm ext})\Psi\,,
\end{equation}
in which $\Phi_{\rm ext}$ \& $\Phi_{\rm self}$ are the gravitational potentials of the bar
or other external perturber and of the condensate itself; $\Phi_{\rm self}$ satisfies
Poisson's equation with the mass density $\rho_{\rm c}=m|\Psi|^2$ of the condensate as
source.  Adopting the usual Jeans swindle, consider perturbations to a background state of
uniform density and vanishing $\Phi$.  The unperturbed wavefunction $\Psi_0$ has a
constant modulus $|\Psi_0|=\sqrt{{\nu}_0}$ but a time-dependent phase, because the
term involving $U_0$ in eq.~\eqref{eq:NLSE} doesn't vanish in the background state.  One
sets $\Psi\to\Psi_0(t)[1+\varepsilon(\bs{r},t)]$ and expands eq.~\eqref{eq:NLSE} to first
order in the real and imaginary parts of $\varepsilon$, treating $\Phi_{\rm ext}$ and
$\Phi_{\rm self}$ as well as $\varepsilon$ as first-order quantities.  In the unforced case
$\Phi_{\rm ext}=0$, the dispersion relation for Fourier modes
$\varepsilon(\bs{r},t)\propto\exp(i\bs{k\cdot r}-i\omega t)$ becomes
\begin{equation}
  \label{eq:dispersion}
  \omega_{\bs{k}}^2= (2U_0 k^2-4\pi Gm){\nu}_0 +\left(\frac{\hbar k^2}{2m}\right)^2.
\end{equation}
As usual with the GP Equation, the last term on the right represents single-particle
exitations; it is small for long-wavelength modes $k\ll \sqrt{2m\cs}/\hbar$, with sound
speed $\cs\equiv\sqrt{2U_0{\nu}_0}$.
If one neglects this term, then eq.~\eqref{eq:dispersion} matches the results
of a Jeans analysis for a classical ideal fluid, $\omega^2=\cs^2 k^2-4\pi G\rho_0$ \citep[e.g.][]{BT}.
If one now introduces a rigid perturbing potential that moves at constant velocity $\bs{V}$
through the condensate, $\Phi_{\rm ext}(\bs{r}-\bs{V}t)$, and writes $\tilde\Phi_{\rm
  ext}(\bs{k})$ for the spatial fourier transform of this potential at any time, then
after transients have decayed, the component of the wave drag along $\bs{V}$ is
\begin{equation}
  \label{eq:drag}
  F_{\rm drag}=-
  \frac{\rho_0}{V}\int\frac{d\bs{k}}{(2\pi)^3}\,\delta(\omega_{\bs{k}}-\bs{k\cdot
    V})|\bs{k}\tilde\Phi_{\rm ext}(\bs{k})|^2.
\end{equation}
Since the fourier components $\tilde\Phi_{\rm ext}(\bs{k})$ are negligible at $k\gtrsim\sqrt{2m\cs}/\hbar$,
this is effectively the same drag as for an ideal fluid with equation of state
$P=K\rho^2$.  The quantum-mechanical nature of the condensate plays no direct role.  
To the extent that the self-gravity of the RDM is slight on the scale of the perturber,
in other words  $\tilde\Phi_{\rm ext}(\bs{k})$ is unimportant for $k^2\lesssim4\pi
G\rho_0/\cs^2$,  the drag vanishes unless $V$ is supersonic.

The formal result \eqref{eq:drag} for the wave drag clearly holds more generally for ideal
fluids in which the dispersion relation may differ from \eqref{eq:dispersion}.  Thus it
should hold even when the RDM has a normal (nondegenerate) component, as it must at finite
temperature.  In a realistic case where the background state is not uniform, however, the
RDM gas would be stratified, i.e. it would have entropy gradient parallel to the
background gravitational field, so that waves restored by buoyancy (internal waves/
g~modes) might be excited at subsonic velocities.  As will be shown in \S\ref{sec:halos},
an isothermal RDM halo at nonzero temperature consists of a core that is almost pure
condensate, and has nearly the $n=1$ Emden profile, surrounded by an extended
nondegenerate ``atmosphere,'' with a sharp shelf in the density profile at the edge of the
core.  Although we have not calculated it, the coupling of a subsonicly rotating bar potential to the
g~modes would probably occur mainly at this shelf, with a drag
proportional to the density in the nondegenerate component there; because that density is
much less than the central density of the degenerate core, the drag on the bar would be
expected to be considerably reduced compared to the estimates of \cite{DS2000} for
collisionless DM at the same overall mass-to-light ratio.

\section{Thermodynamics}\label{sec:thermo}

Apart from the above clarification of its implications for dynamical friction, our main
progress concerning RDM since \cite{G2000} has been to construct isothermal spherical halo
models, which has lead us to a lower bound on the collisionality of the RDM that is, while
somwhat imprecise, nevertheless orders of magnitude less than what has been inferred from
the Bullet Cluster---hence ruling out RDM, we think.

For this purpose we needed to derive the equation of state of RDM at finite temperature,
which we did by calculating the partition function of RDM considered as an almost ideal
gas.  ``Almost'' means that the scattering length $\lambda_{\rm scatt}\equiv mU_0/\hbar^2$
($\approx\sqrt{\pi\sigma_{\rm scatt}}$) is much less than the de~Broglie wavelength
$\lambda_{\rm dB}\equiv h/\sqrt{2mkT}$.  Then, just as in a classical dilute gas, the
(non-condensate) particles can be regarded as belonging to well-defined momentum states
between collisions.  On the other hand, $\lambda_{\rm dB}$ should be small compared to
galactic scales so that we may consider the partition function of a homogenous boxful of
RDM.

The almost-ideal boson gas is a well-studied physical model for a superfluid.  The
standard treatment in textbooks \citep[e.g.][]{PL} emphasizes the dynamics of
quasi-particle excitations when most of the particles are in the condensate, and is
couched in terms of annihilation and creation operators for these excitations.  Our
approach is cruder but, we hope, sufficient to determine the pressure in local
thermodynamic equlibrium (LTE).  Insofar as possible, we follow elementary treatments of a
noninteracting ideal boson gas, describing microstates by the numbers of quanta
$(n_0,n_1,n_2,\ldots,n_i,\ldots)\equiv\vec n$ in each single-particle momentum state
$\ket{\bs{p}_i}$, these states being discrete if the gas is imagined to be confined to a
finite volume $V$, and ordered by their kinetic energies so that $p_i^2\le p_j^2$ if $i\le
j$.  The total number of particles $N=\sum_i n_i$, and the number density
${\nu}=N/V$.  We presume that there is a unique state $\ket{\bs{p}_0}$ of zero
kinetic energy (the ground state).  The condensate, if present, will be characterized by
the occupation number $n_0$ of the ground single-particle state being macroscopic,
i.e. comparable to $N$.  The kinetic energy $|\bs{p}_i|^2/2m$ of all other states is
higher than that of the ground state by at least $O(\hbar^2/2mV^{2/3})$; while this is
small for macroscopic volume $V$, it is larger than $\kB T/N$ by a factor that is
$O(N^{1/3})$ at fixed ${\nu}=N/V$.  Therefore, a macroscopic occupation number in
any state other than $\ket{\bs{p}_0}$ is always exponentially suppressed.

The energy of a general microstate $\vec n$ is
\begin{align}\label{eq:Esums}
  E(\vec n,V) 
&= \sum_{i\ge0}\frac{p_i^2}{2m}n_i + 
 U_0\frac{N^2}{2V}+\frac{U_0}{2V}\underset{{\scriptstyle i\ne j}}{\sum\sum} n_i n_j\nonumber\\
&= \sum_{i\ge0}\frac{p_i^2}{2m}n_i + \frac{U_0}{V}\left[N^2-\tfrac{1}{2}\sum_i n_i^2\right]\nonumber\\
&\approx \sum_{i\ge0}\frac{p_i^2}{2m}n_i + \frac{U_0}{2V}\left[2N^2-n_0^2\right].
\end{align}
The double sum in the first line represents the exchange energy for pairs in
distinct states, and is rewritten as $(\sum_i n_i)^2-\sum_i n_i^2$ in the second line.
Now
\begin{align*}
  \frac{U_0}{2V}\sum_{i>0}n_i^2\le   \frac{U_0}{2V}\,\max\limits_{j>0} n_j\sum_{i>0}n_i\le
   \tfrac{1}{2}{\nu}U_0\,\max\limits_{j>0} n_j.
\end{align*}
Following the remarks in the last paragraph, $n_j$ is microscopic for all $j>0$.  So the
error in the last line of \eqref{eq:Esums} grows more slowly with $N$ than $E(\vec n,V)$
as $N\to\infty$ at constant $N/V$ and therefore can be neglected in the thermodynamic
limit.  This step is essential, because the negative sign in front of $\sum_i n_i^2$ would
otherwise prevent us from summing independently over all the $n_i$ with a chemical
potential $\mu$ to enforce $\bar N=N$, as one does to evaluate the grand-canonical
partition function of a \emph{noninteracting} gas, because the sums would diverge.  But
with the approximation \eqref{eq:Esums}, one can eliminate $N$ and $n_0$ in favor of the
intensive variables $\nu_0(\mu,t)\equiv \bar n_0/V$ and $\nu(\mu,T)\equiv\bar N/V$,
leaving $V$ as the sole extensive variable, and thereby obtain the grand-canonical
partition function of the interacting system, and hence the pressure.  Here $\nu_0\le\nu$
is the number density of condensate particles only, not to be confused with the number
density of the background state in the wave-drag calculation of the previous section.

\begin{figure}[ht]
\begin{center}
\includegraphics[width=0.3\columnwidth]{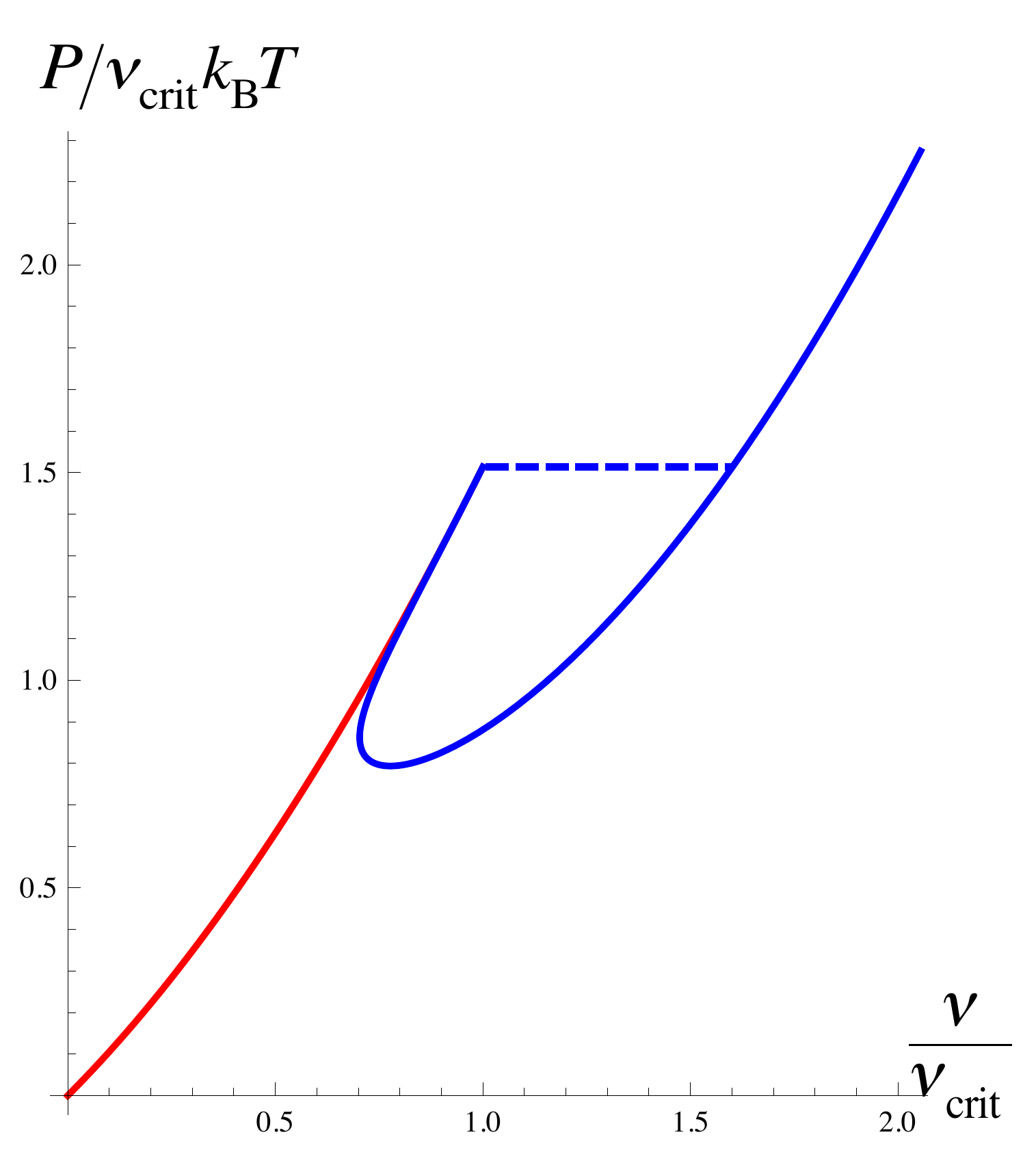}
\caption{Scaled pressure $\hat P$ versus scaled number density
  $\hat\nu$ for interaction strength $\theta=1$ [eq.~\eqref{eq:thetadef}].}
\label{fig:Pvsnu}
\end{center}
\end{figure} 
 
One finds that the condensate appears when the number density is greater than $\nu_{\rm
  crit}(T)=\zeta(3/2)(m\kB T/h\hbar)^{3/2}$, which is the same as thecritical density as
for a noninteracting gas.  In terms of scaled number densities $\hat\nu\equiv\nu/\nu_{\rm
  crit}$ and $\hat\nu_0\equiv\nu_0/\nu_{\rm crit}$, a scaled pressure $\hat P\equiv
P/\nu_{\rm crit}\kB T\propto P/T^{5/2}$, and a dimensionless measure of the repulsion term
\begin{equation}
  \label{eq:thetadef}
  \theta\equiv \frac{U_0}{\kB T}\nu_{\rm crit} \
  =\tfrac{1}{2}\zeta(3/2)\,\frac{\sigma_{\rm scatt}^{1/2}}{\lambda_{\rm dB}}\,,
\end{equation}
the equation of state is
\begin{subequations}\label{eq:eos}
  \begin{align}
    \label{eq:Phat}
    \hat P &= \left(\hat\nu^2-\tfrac{1}{2}\hat\nu_0^2\right)\theta + \left.\Li_{5/2}(z)\right/\zeta(3/2),\\
    \label{eq:nuhat}
    \hat\nu &= \left.\Li_{3/2}(z)\right/\zeta(3/2)\ +\max(\hat\nu_0,0),\\
    \label{eq:zofnu0}
    z &= \exp(-\theta\hat\nu_0)\quad\mbox{if and only if}\quad\hat\nu_0>0.
  \end{align}
\end{subequations}
Here $\Li_s(z)$ is the polylogarithm [$\Li_s(z)=\sum_{k=0}^\infty z^k/k^s$ where this
converges], which relates to the Bose-Einstein integrals when $s$ is an integer or
half-integer.  The parameter $z$ is a sort of fugacity.  When $\hat\nu_0=0$, meaning that
the condensate is absent, then $z$ is an implicit function of $\hat\nu$ defined by
eq.~\eqref{eq:Phat}; otherwise, eq.~\eqref{eq:zofnu0} gives $z$ explicitly.  The
noninteracting gas is recovered by setting $\theta=0$: then pressure is constant once the
condensate appears, if temperature is fixed.

A plot of the equation of state \eqref{eq:eos} is shown in Fig.~\ref{fig:Pvsnu} for
$\theta=1$, corresponding to strong scattering, although the steps leading to
eqs.~\eqref{eq:eos} probably cannot be justified unless $\theta\ll 1$.  The condensate is
absent along the red section, and present along the blue.  The part of the curve beneath
the dashed line segment is unphysical, and the endpoints of this segment define two phases
in contact; it can be shown that $\oint\hat P\ud\hat\nu=0$ around the loop defined by the
dashed segment and the unphysical lobe below it.  The density jump at the phase transition
scales $\propto\theta$ when $\theta\ll1$.  In that limit, however, then because of the
factors of $\theta$ in eqs.~\eqref{eq:nuhat} and \eqref{eq:zofnu0}, the pressure rises
only slowly with increasing $\nu>\nu_{\rm crit}$ until $\hat\nu\gtrsim\theta^{-1}$, at
which point $\nu\approx\nu_0$ because $z$ and the polylogarithms become small.

In other words, while the actual phase transition is rather weak for weak collisionality
($\sigma_{\rm scatt}\ll \lambda_{\rm dB}^2$), in the same limit an isothermal system goes from
non-degenerate to almost entirely degenerate over a small increase in pressure but a large
increase in density.  It is this feature of the equation of state that leads to constraints
on the collisionality of RDM from galactic rotation curves.

\section{RDM halos}\label{sec:halos}

The simplest model for dark halo in virial equilibrium that resembles inferences from
observations of galaxies is an isothermal sphere.  For a classical ideal gas or system of
collisionless WIMPs, the isothermal sphere has a unique density profile, apart from
rescalings of the central density and core radius.  It is the limit of the sequence of
Emden spheres, as is also discussed in
Chandrasekhar's first book \cite{Chandra_SS}, and is characterized by constant
velocity dispersion and constant $P/\rho$ at all radii.  Although the density profile cannot be
obtained in closed form, $\rho(r)\sim r^{-2}$ at $r\gg r_{\rm c}$; this implies a mass
contained with radius $r$ that is asymptotically linear, $M(r)\sim r$, and therefore a
rotation velocity for circular orbits, $v_{\rm c}(r)=GM(r)/r$, that is asymptotically
constant, in general agreement with rotation curves of spiral galaxies out to the largest
observable radii.  The total mass profile (DM+baryons) appears to be remarkably
close to $\rho\propto r^{-2}$ within at least the luminous regions of elliptical galaxies
\cite{Koopmans09}.

\begin{figure}[ht]
  \centering
  \includegraphics[width= 0.4\columnwidth]{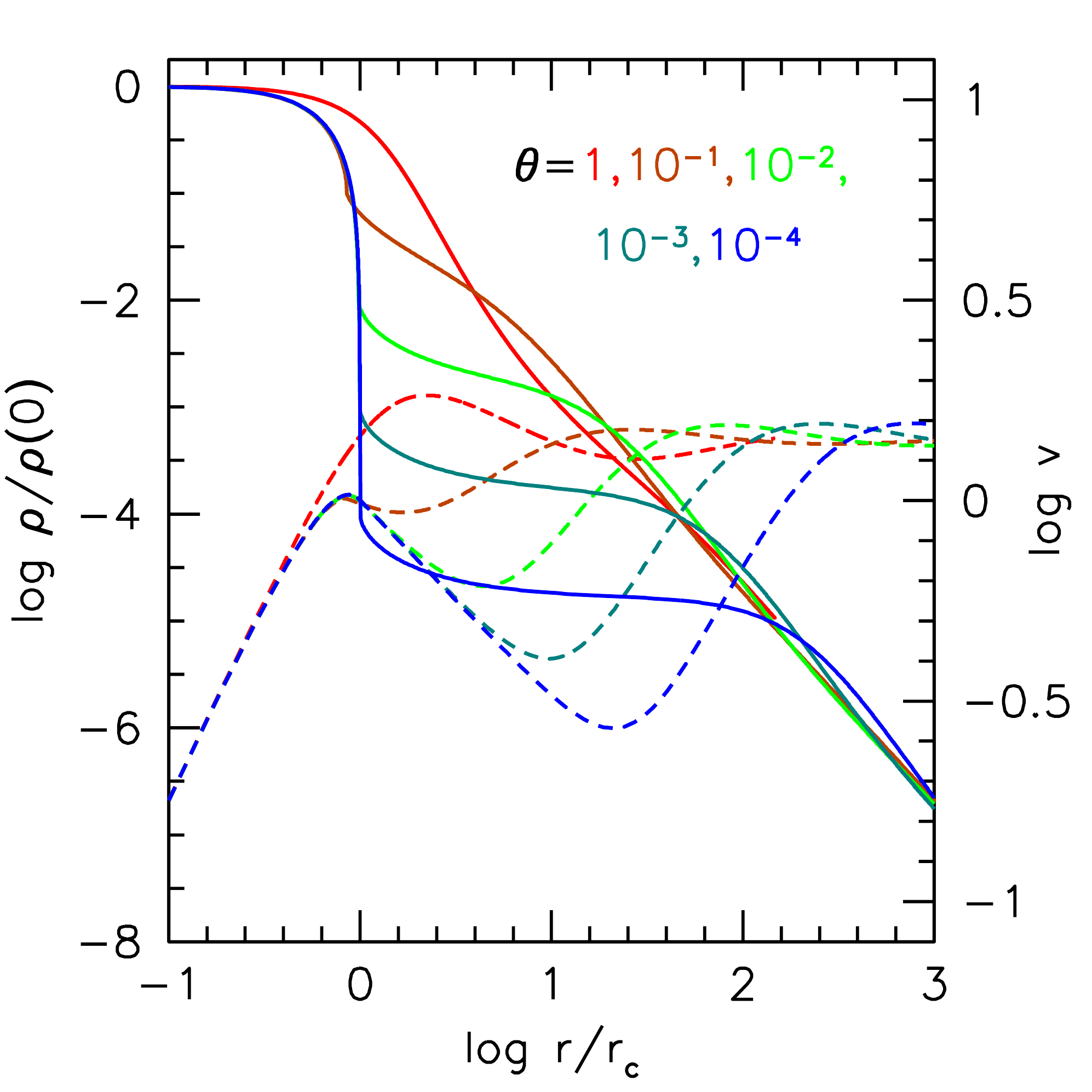}
  \caption{Self-gravitating isothermal spheres of RDM for various collisionalities
    $\theta$ [eq.~\eqref{eq:thetadef}].
    {\it Solid curves and left axis:} mass density.  {\it Dashed curves and righthand axis:} rotation curves.}
  \label{fig:halos}
  \end{figure}

  Using the equation of state \eqref{eq:eos}, one can construct self-gravitating
  isothermal spheres of RDM.  The condensate dominates within the core radius $r_{\rm
    c}\equiv \pi a$, and the density profile there is approximately the same as for zero
  temperature, namely an $n=1$ Emden model.  At much larger radii, $\rho\propto r^{-2}$ as
  for the classical isothermal sphere.  The behavior at intermediate radii is quite
  different from the classical case and depends strongly upon the collisionality parameter
  \eqref{eq:thetadef}.  This is shown in Figure~\ref{fig:halos}.  All of these models are
  scaled to the same central density, core radius, and asymptotic rotation velocity, which
  turns out to require that the scaled central number density
  $\hat\nu(0)\propto\theta^{-1}$; so apart from dimensional scalings, they form a
  one-parameter sequence.  Paradoxically, the strongly collisional model, $\theta=1$,
  looks most like the classical case, but for $\theta\ll 1$, the density drops sharply at
  the edge of the core and is approximately constant out to $r\approx\theta^{-1/2} r_{\rm
    c}$.  The explanation for this behavior is that the doubling of the repulsion due to
  exchange forces acting on the nondegenerate component within the core (the nondegenerate
  component is never entirely absent) renders this component almost unbound, so that its
  density becomes small and declines slowly until such a radius that its own interior mass
  begins to be larger than that of the core.  This structure is analogous to that of a red
  giant, in which the central parts are supported by degeneracy pressure and there is a
  large increase in entropy across the hydrogen burning shell, leading to a distended
  convective envelope.

The drop in density at the edge of the core is associated with a drop in the rotation
curve, which eventually recovers and is approximately constant beyond $\theta^{-1/2}r_{\rm c}$.
Fig.~\ref{fig:halos} shows that $v_{\rm c}$ drops by a factor $\sim 2$ for
$\theta=10^{-4}$.  A drop this large or larger would seem to be in contradiction with what
is inferred for the contribution of the halo to $v_{\rm c}$ after that of the baryonic
matter is subtracted.  Thus, insofar as RDM halos are (or would be) isothermal, we can
safely conclude that $\theta\ge 10^{-4}$.

\section{Limits on the RDM boson mass}\label{sec:mass}

Eliminating $U_0$ between $\sigma_{\rm scattt}=(mU_0)^2/\pi\hbar^4$ and $r_{\rm c}=\pi
a=\sqrt{\pi U_0/4Gm^2}$ leads to $\sigma = (4/\pi^2)G^2\hbar^{-4}m^6 r_{\rm c}^4$.
Therefore, one can derive an upper bound on $m$ in terms of the constraint inferred by
\cite{Randall_etal2008} for $\sigma/m$ from the Bullet Cluster:
\begin{subequations}\label{eq:bounds}
\begin{equation}
  \label{eq:mmax}
m< 7\times 10^{-4}\left(\frac{\sigma/m}{1.25{\rm\,cm^2\,g^{-1}}}\right)^{1/5}
\left(\frac{r_{\rm c}}{1\,{\rm kpc}}\right)^{-4/5}\left.{\rm eV}\right/c^2.
\end{equation}
On the other hand, taking $\lambda_{\rm dB}\approx h/mv_{\rm c}$ in
eq.~\eqref{eq:thetadef}, where $v_{\rm c}$ is the asymptotic circular velocity of the
halo, one has a lower bound on $m$ in terms of the constraint $\theta>10^{-4}$ obtained in
the last section:
\begin{equation}
  \label{eq:mmin}
  m > 9. \left(10^4\theta\right)^{1/4}\left(\frac{v_{\rm c}}{100\,{\rm km\,s^{-1}}}\right)^{-1/4}
\left(\frac{r_{\rm c}}{1\,{\rm kpc}}\right)^{-1/2}\left.{\rm eV}\right/c^2.
\end{equation}
This corresponds to $\lambda_{\rm dB}\lesssim 0.05{\rm\,cm}$ for $v_{\rm c}=100\,{\rm km\,s^{-1}}$.
\end{subequations}

The two inequalities \eqref{eq:mmax} \& \eqref{eq:mmin} are clearly incompatible for any
astronomically acceptable values of core radius and circular velocity.  Therefore, it
seems that repulsive dark matter, at least of the sort we have been considering here, can
be ruled out.

The lower bound \eqref{eq:mmin} derives from our assumption that the halo is isothermal,
however, so one should ask whether that assumption is reasonable.  \emph{Local}
thermodynamic equilibrium should be established on a collision time $t_{\rm coll}\sim
\lambda_{\rm mfp}/v_{\rm c}$, but global isothermality is achieved on a conduction time
$t_{\rm cond}\sim r^2/(v_{\rm c}\lambda_{\rm mfp})$.  If the mean-free path were small
compared to galactic scales, one could have $t_{\rm cond}\gg t_{\rm coll}$.  As noted in
our discussion earlier, however, $\sigma/m\approx 1\,{\rm g\,cm^{-1}}$ corresponds to
$\lambda_{\rm mfp}\sim r$ in the solar neighborhood, so if eq.~\eqref{eq:mmax} were an
approximate equality, then these timescales would be comparable to one another, and rather
less than a Hubble time.  On the other hand, if the mass were very much less than
\eqref{eq:mmax}, then the RDM would suffer little scattering, so that even $t_{\rm coll}$
could be longer than the age of the Galaxy.  Thermodynamics considerations would not
apply.  And yet, despite the lack of scattering, the repulsion could still provide a
minimal $r_{\rm c}\sim 1\,{\rm kpc}$ since (as we have already noted) the core radius and
the scattering cross section involve different combinations of $U_0$ and $m$.

\acknowledgments
JG would like to thank the organizers and sponsors of the Chandrasekhar Centenary
Conference for a most enjoyable and stimulating meeting, and the members and staff of the
Indian Institute of Astrophysics.

\bibliographystyle{pramana}
\bibliography{goodmanchandra}

\end{document}